\documentclass[10pt]{article}

\usepackage{amscd,amsmath,amssymb,latexsym,stmaryrd,bbm}

\catcode`@=11
\renewcommand{\section}{\@startsection{section}{1}{0pt}{\medskipamount}
{\medskipamount}{\large\bf}}
\numberwithin{equation}{section}
\catcode`@=12

\newcommand{\p}[1]{(\ref{#1})}

\newcommand{\bpsi}{{\bar\psi}{}}
\newcommand{\eps}{\varepsilon}

\newcommand{\R}{\mathbb R}

\newcommand{\unity}{\mathbbm{1}}

\newcommand{\be}{\begin{equation}}
\newcommand{\ee}{\end{equation}}
\newcommand{\bea}{\begin{eqnarray}}
\newcommand{\eea}{\end{eqnarray}}
\newcommand{\ba}{\begin{array}} 
\newcommand{\ea}{\end{array}}

\def\pa{\partial}
\def\diff{{\textrm{d}}}
\def\sfrac#1#2{{\textstyle\frac{#1}{#2}}}
\def\={\ =\ }
\def\und{\qquad\textrm{and}\qquad}
\def\im{{\rm i}}
\def\ep{{\rm e}}
\def\ph{\phantom{-}}

\begin{document}

\begin{titlepage}
\setcounter{page}{0}
\begin{flushright}
ITP--UH--18/10
\end{flushright}

\vskip 2.0cm

\begin{center}

{\LARGE\bf
Many-particle mechanics \\[8pt]
with $D(2,1;\alpha)$ superconformal symmetry}

\vspace{12mm}

{\Large Sergey Krivonos$\,{}^*$ \ and \ \Large Olaf Lechtenfeld$\,{}^+$}
\\[8mm]
\noindent ${}^*${\em Bogoliubov Laboratory of Theoretical Physics, JINR \\
141980 Dubna, Russia }
\\[8mm]
\noindent ${}^+${\em Institut f\"ur Theoretische Physik,
Leibniz Universit\"at Hannover \\
Appelstrasse 2, 30167 Hannover, Germany }

\vspace{12mm}

\begin{abstract}
\noindent
We overcome the barrier of constructing ${\cal N}{=}4$ superconformal
models in one space dimension for more than three particles.
The $D(2,1;\alpha)$ superalgebra of our systems is realized on the
coordinates and momenta of the particles, their superpartners and one complex 
pair of harmonic variables. The models are determined by two prepotentials,
$F$ and~$U$, which must obey the WDVV and a Killing-type equation plus
homogeneity conditions. We investigate permutation-symmetric solutions,
with and without translation invariance. Models based on deformed $A_n$ and
$BCD_n$ root systems are constructed for any value of~$\alpha$, and
exceptional $F_n$-type and super root systems admit solutions as well.
Translation-invariant mechanics occurs for any number of particles 
at $\alpha{=}{-}\sfrac12$ ($osp(4|2)$ invariance as a degenerate limit)
and for four particles at arbitrary~$\alpha$ (three series).
\end{abstract}

\end{center}
\end{titlepage}

\section{Introduction}

\noindent
It has proved surprisingly difficult to construct ${\cal N}{=}4$ superconformal
mechanics for more than three particles~\cite{tolik}--\cite{lst}.\footnote{
For recent results on three-particle systems, see~\cite{bel1}.
In principle, one may add also harmonic potentials~\cite{bel2}.}
The Hamiltonian (or the action) of such models is determined by two
scalar prepotentials, $F$ and~$U$, which are functions of the bosonic
particle coordinates~$\{x^i\}$ and obey two nonlinear differential equations,
namely the celebrated WDVV~equation for~$F$ and a Killing-type equation
for~$U$ in the $F$~background. On top of this, conformal invariance
imposes some homogeneity conditions on $U'$ and $F'''$. Each solution
to all equations produces a consistent many-particle model.

In the one dimension (time) of mechanical systems, four supercharges
implies invariance under the exceptional superalgebra $D(2,1;\alpha)$, 
for some value of the real parameter~$\alpha$.\footnote{
Permuting the three $sl(2)$ subalgebras of $D(2,1;\alpha)$ relates
$\alpha\leftrightarrow-1{-}\alpha\leftrightarrow\sfrac1\alpha\leftrightarrow
\sfrac{-1-\alpha}{\alpha}\leftrightarrow\sfrac{-1}{1+\alpha}\leftrightarrow
\sfrac{-\alpha}{1+\alpha}$~\cite{fss}.}
For the special cases of $D(2,1;0)\simeq su(1,1|2)\niplus su(2)$
and $D(2,1;1)\simeq D(2,1;-\sfrac12)\simeq osp(4|2)$ some results were
obtained~\cite{fil1,klp}. 
On the one hand, by gauging the U($n$) isometry of matrix superfield models, 
one can construct U(2) spin-extended mechanics for arbitrary values 
of~$\alpha$~\cite{fil1,fil3}. However, the particle coordinates parametrize a 
non-flat target space, except for $\alpha{=}-\sfrac12$, 
i.e.~the $osp(4|2)$ case.
On the other hand, for $\alpha{=}0$ a superspace approach produced an
alternative formulation, which allowed for the construction of a few 
nontrivial four-particle solutions~\cite{klp}. In this case,
the `structure equations' (writing $W$ instead of~$U$ here) can be cast 
in the form~\cite{lst}
\be
\widehat{F}\wedge\widehat{F}\=0 \und
(\widehat{\diff}-\widehat{F})\,|W\rangle \=0\ ,
\ee
with the shorthand notation ($i,j,k,\ldots$ label the particles)
\be
\widehat{F} = \diff x^k\widehat{F}_k = \bigl(\diff x^k F_{ijk}\bigr) \ ,\qquad
\widehat\diff = \diff x^k\widehat\pa_k = \bigl(\diff x^k\pa_k\delta_{ij}\bigr)
\und |W\rangle = \bigl(W_i\bigr)
\ee
packaging the derivatives of $F$ and~$W$ in a matrix-valued one-form and
a (ket) vector, respectively.
The question remains whether it is possible to construct
$D(2,1;\alpha)$ invariant models with more than three particles for other
values of~$\alpha$ and perhaps without a spin extension.

In this paper we answer this question in the affirmative.
By introducing just a single set of bosonic spin variables
$\{u^a,\bar{u}_a|\,a{=}1,2\}$ with Poisson brackets
\be
\bigl\{u^a,\bar{u}_b\bigr\}=-\im\,\delta^a_{\ b}
\qquad\textrm{and}\qquad\textrm{$su(2)$ currents}\quad
J^{ab}=\sfrac{\im}{2}\left( u^a {\bar u}^b+u^b{\bar u}^a \right)\ ,
\ee
we slightly generalize the ansatz of~\cite{glp2} for the supersymmetry 
generators. As a consequence, the structure equations (for any~$\alpha$) 
get modified to
\be \label{str}
\widehat{F}\wedge\widehat{F}\=0 \und
(\widehat{\diff}-\widehat{F})\,|U\rangle \= (\widehat{\diff}U)\,|U\rangle\ ,
\ee
where $|U\rangle=\bigl(U_i\bigr)$ is distinguished from $|W\rangle$.
The WDVV equation is unchanged, and the integrability of the Killing-type 
equation still yields $(\widehat{F}\wedge\widehat{F})|U\rangle=0$.
Despite being nonlinear, the new term on the right-hand side is not
a nuisance but actually a benefit, 
because it greatly enhances the solvability of the equation!
In addition, one still has ($\alpha$-dependent) homogeneity conditions
for $F$ and~$U$. The new spin variables appear in the Hamiltonian merely
via its $su(2)$ currents $J^{ab}$.

Since we aim at describing a collection of identical particles,
we are not interested in arbitrary solutions of~(\ref{str}), but only those
which are invariant under permutations of the particle labels.
We do allow for translation non-invariance, however, because of the
canonical relation between a translation-invariant system of $n{+}1$
particles to the reduced $n$-dimensional system of their relative coordinates,
after decoupling the center of mass.
In the next section, we derive the generic formulae for our new models, 
present the universal ansatz for $F$ and~$U$ in terms of a collection of 
(co)vectors and their orbits under the permutation group and outline our
strategy for solving~(\ref{str}). The following four sections present
our explicit solutions $(F,U)$ for the $A$-type, $BCD$-type, $EF$-type
and non-Coxeter-type series of known WDVV configurations~$F$. There exist 
families of solutions as well as sporadic ones. Finally, we conclude with 
a summary and some observations. All irreducible four-particle solutions
are collected in an Appendix.

\newpage

\section{$D(2,1;\alpha)$ invariant many-particle system}

\noindent
We consider $n{+}1$ particles on a real line, 
with (bosonic) coordinates and momenta $\{x_i,p_i|\,i{=}1,\ldots,n{+}1\}$ 
as well as associated complex pairs of fermionic variables 
$\{\psi^a_i,\bpsi_{ai}|\,i{=}1,\ldots,n{+}1,\,a{=}1,2\}$.\footnote{
Viewed as a one-particle system, the bosonic target is $\R^{n+1}$.
Its metric~$(\delta_{ij})$ allows us to pull down all particle indices.\\
Spinor indices are raised and lowered with the invariant tensor
$\eps^{ab}$ and its inverse $\eps_{ba}$, respectively.}
In addition, we introduce one set of (bosonic) spin variables 
$\{u^a,\bar{u}_a|\,a{=}1,2\}$ parametrizing an internal two-sphere.
The basic Poisson brackets read
\be\label{PB}
\bigl\{x_i, p_j\bigr\}=\delta_{ij}\ ,\qquad 
\bigl\{\psi^a_i,\bpsi_{bj}\bigr\}=\sfrac\im 2\delta^a_{\ b}\delta_{ij}\ ,\qquad
\bigl\{u^a,\bar{u}_b\bigr\}=-\im\,\delta^a_{\ b}\ .
\ee
We would like to realize the ${\cal N}{=}4$ superconformal algebra
$D(2,1;\alpha)$ on the (classical) phase space of this mechanical system,
thereby severely restricting the particle interactions.
It is convenient to start with an ansatz for the supercharges 
$Q^a$ and $\bar Q_a$. Previously~\cite{glp2}, they were chosen 
(in our normalization) as
\be\label{oldQ}
Q^a=p_i\psi^a_i+\im W_i(x)\psi^a_i+\im F_{ijk}(x)\psi^b_i\psi_{bj}\bpsi^a_k
\und \bar Q_a=
p_i\bpsi_{ai}-\im W_i(x)\bpsi_{ai}+\im F_{ijk}(x)\bpsi_{bi}\bpsi^b_j\psi_{ak}
\ee
with $F_{ijk}$ being totally symmetric.
This ansatz was successful for the algebra 
$D(2,1;0)\simeq su(1,1|2)\niplus su(2)$ with a central charge~$C$,
upon solving some integrability conditions for $W_i$ and $F_{ijk}$
including the celebrated WDVV equation~\cite{w,dvv}. However, it turned out
to be very hard to generate explicit solutions for $W_i$ beyond three
particles~\cite{klp}.

Here, we utilize the spin variables to slightly generalize this ansatz to
\be\label{Q}
Q^a=p_i\psi^a_i+U_i(x)J^{ab}\psi_{bi}+\im F_{ijk}(x)\psi^b_i\psi_{bj}\bpsi^a_k
\und {\bar Q}_a= 
p_i\bpsi_{ai}-U_i(x)J_{ab}\bpsi^b_i+\im F_{ijk}(x)\bpsi_{bi}\bpsi^b_j\psi_{ak}
\ee
with the $su(2)$ currents
\be\label{J}
J^{ab}=\sfrac{\im}{2}\left( u^a {\bar u}^b+u^b{\bar u}^a \right)
\qquad \Rightarrow \qquad 
\left\{J^{ab},J^{cd}\right\}=-\epsilon^{ac}J^{bd}-\epsilon^{bd}J^{ac}\ .
\ee
The spin variables just serve to produce these currents and do not appear 
by themselves. 

Let us try to build the $D(2,1;\alpha)$ algebra based on \p{Q}.
Firstly, the ${\cal N}{=}4$ super-Poincar\'e subalgebra
\be\label{Poincare}
\left\{ Q^a, Q^b\right\} =0 \und 
\left\{ Q^a, {\bar Q}_b \right\} =2\im\,\delta^a_{\ b} H
\ee
defines a Hamiltonian~$H$ and enforces the following conditions on
our functions $U_i$ and $F_{ijk}$,
\bea
&& \pa_i U_j-\partial_j U_i =0\ ,\quad \pa_i F_{jkl}-\pa_j F_{ikl}=0\ , 
\label{integr} \\[4pt]
&& F_{kim}F_{mj\ell} -F_{kjm}F_{mi\ell}=0\ , 
\label{wdvv} \\[4pt]
&& -\partial_i U_j +U_i U_j +F_{ijk}U_k=0\ . 
\label{eq1}
\eea
The integrability conditions \p{integr} are solved by
\be\label{integr1}
U_i =\pa_i U \und F_{ijk} = \pa_i \pa_j \pa_k F
\ee
with two scalar prepotentials $F(x)$ and $U(x)$,
and hence we read subscripts on $U$ and $F$ as derivatives.\footnote{
Note that $U(x)$ and $F(x)$ are defined only up to polynomials of degree zero
and two, respectively.}
Thus, the other two conditions become nonlinear differential equations
for $F(x)$ and $U(x)$, whose solutions define the various possible models.
With the above conditions fulfilled, the Hamiltonian acquires the form
\be\label{Ham}
H\=\sfrac14 p_i p_i +\sfrac18 J^{ab}J_{ab}\,U_i U_i
-\im U_{ij}\,J^{ab}\,\psi_{ai}\bpsi_{bj}-
\sfrac12 F_{ijk\ell}\,\psi^a_i\psi_{aj}\,\bpsi_{bk}\bpsi^b_\ell\ .
\ee
One may check that $[H,J^{ab}J_{ab}]=0$, and thus the Casimir 
$J^{ab}J_{ab}=:g^2$ appears as a coupling constant in the bosonic potential
\be
V \= \sfrac{g^2}{8}\,U_iU_i\ .
\ee

Secondly, for the full $D(1,2;\alpha)$ superconformal invariance one has 
to realize the additional generators. This can be done via
\be\label{Dalpha}
D= -\sfrac12 x_i p_i\ ,\qquad K= x_i x_i\ ,\qquad  
S^a = - 2 x_i \psi^a_i\ ,\qquad {\bar S}_a = - 2 x_i \bpsi_{ai}\ ,
\ee
together with two sets of composite $su(2)$ currents,
\be\label{su2}
{\cal J}^{ab} = J^{ab} +2\im \psi_i^{(a}\bpsi_i^{b)} \und
I^{11}=\im \psi^a_i\psi_{ai}\ ,\quad 
I^{22}=-\im\bpsi_{ai}\bpsi^a_i\ ,\quad 
I^{12}=\im \psi^a_i \bpsi_{ai}\ ,
\ee
in the notation of \cite{fil3}.
Now, dilatation invariance requires homogeneity,
\be\label{dil}
(x_i\pa_i+1)\,U_j=\pa_j(x_i U_i)=0 \und 
(x_i\pa_i+1)\,F_{jk\ell} =\pa_j(x_i F_{ik\ell})=0\ .
\ee
Thirdly, the remaining superalgebra commutators only fix the integration 
constants to
\be\label{conD}
x_i U_i =2\alpha \und x_i F_{ijk}=-(1{+}2\alpha)\,\delta_{jk} 
\qquad\Rightarrow\qquad (x_i\pa_i-2)\,F=-\sfrac12(1{+}2\alpha)\,x_ix_i\ .
\ee

It is instructive to compare our equations with the ones obtained 
in~\cite{glp2} with the ansatz~\p{oldQ} for the case of~$\alpha{=}0$. 
The integrability condition~\p{integr} and the WDVV equation~\p{wdvv}
emerged there as well, but the Killing-type equation lacked the term
quadratic in~$U$. Still, we may map our equation to theirs by defining
\be \label{Weq}
W = \ep^{-U} \qquad\Rightarrow\qquad
W_{ij}-F_{ijk}W_k=0\ .
\ee
The inhomogeneities of $U$ and $W$ are also related: 
At $\alpha{=}0$ we may introduce a central charge~$C$ 
by extending the first equation of~\p{conD} to
\be
x_i U_i = C\,\ep^U \qquad\Leftrightarrow\qquad x_i W_i=-C\ .
\ee
Here, $U\equiv0$ is an option via $C{=}0$, but not so for $\alpha{\neq}0$.

For $\alpha{\neq}0$, no central charge is allowed, and the prepotentials
take the form
\be
U(x)\ \sim\ \alpha\,\ln x^2\ +\ U_0(x) \und 
F(x)\ \sim\ -\sfrac14\,(1{+}2\alpha)\,x^2\ln x^2\ +\ F_0(x)\ ,
\ee
where $U_0$ and $F_0$ are homogeneous of degree~$0$ and~$2$, respectively.
Clearly, the prepotentials~$F$ for any two values of~$\alpha$ are related
by a mere rescaling as long as $\alpha{\neq}{-}\sfrac12$.
The mathematical literature usually does not introduce a euclidean metric
$\delta_{jk}$ but defines an induced metric $G_{jk}=-x_iF_{ijk}$
which is constant and nondegenerate.
Hence, for any $\alpha{\neq}{-}\sfrac12$ we can import all known 
WDVV solutions~\cite{margra}--\cite{feives2} up to constant coordinate 
transformations. The special case
of $D(2,1;{-}\sfrac12)\simeq D(2,1;1)\simeq osp(4|2)$ only appears
as a singular limit, where $F$ can no longer be `normalized' via~\p{conD}
and the induced metric degenerates.

A global $SO(n{+}1)$ coordinate transformation does not change the structure
of a given model but its physical interpretation, since $x_i$ denote the
particle locations. For a system of identical particles in the absence of
an external potential we should also demand invariance under permutations
of the $x_i$ as well as global translation invariance, $x_i\to x_i+\xi$.
The latter is related to the decoupling of the (free) center-of-mass motion.
Introducing center-of-mass and relative coordinates
\be
X={\textstyle\sum_i}x_i=:\rho\cdot x \und x_i^\perp=x_i-\sfrac1{n+1}X
\qquad\textrm{so that}\quad {\textstyle\sum_i}x_i^\perp=0\ ,
\ee
we can project out the center-of-mass degree of freedom with
\be
P^\parallel=\sfrac1{n+1}\,\rho\otimes\rho \und
P^\perp=\sfrac1{n+1}\left(\begin{smallmatrix}
 n & -1 & \ldots & -1 \\
-1 &  n & \ldots & -1 \\
\vdots & \vdots & \ddots & \vdots \\
-1 & -1 & \ldots & n \end{smallmatrix}\right)\ .
\ee
One finds that 
\be
U(x)= U^\perp(x^\perp) \und
F(x)= F^\parallel(X)+F^\perp(x^\perp) \qquad\textrm{with}\qquad
F^\parallel=-\sfrac14\sfrac{1+2\alpha}{n+1}\,X^2\ln X^2\ ,
\ee
and our equations (\ref{integr})--(\ref{eq1}) 
are also valid for $U^\perp$ and $F^\perp$, while (\ref{conD}) projects to 
\be
x_iU_i^\perp=2\alpha \und
x_iF_{ijk}^\perp=-(1{+}2\alpha)\,P_{jk}^\perp
\qquad\Rightarrow\qquad
(x_i\pa_i-2)\,F^\perp=-\sfrac12(1{+}2\alpha)\,x\,P^\perp x\ ,
\ee
because $\sum_iU_i^\perp=0$ and $\sum_iF_{ijk}^\perp=0$.

However, the set $\{x_i^\perp\}$ is linearly dependent. In order to select
$n$ independent relative coordinates, one should apply an $SO(n{+}1)$ 
transformation which rotates $\rho=(1,1,\ldots,1,1)$ to 
$(0,0,\ldots,0,\sqrt{n{+}1})$.
A possible (but by no means unique) choice for the resulting relative 
coordinates $y=\{y_i|\,i=1,\ldots,n\}$ is
\be \label{embed}
\left(\begin{matrix} 
y_1 \\[6pt] y_2 \\[4pt] \vdots \\[4pt] y_n \\[6pt] y_0 \end{matrix}\right)\=
\left(\begin{matrix}
\sfrac1{\sqrt{1\cdot2}} & \sfrac{-1}{\sqrt{1\cdot2}} & 0 & \ldots & 0 \\[6pt]
\sfrac1{\sqrt{2\cdot3}} & \sfrac1{\sqrt{2\cdot3}} & \sfrac{-2}{\sqrt{2\cdot3}}
& \ldots & 0 \\[4pt]
\vdots & \vdots & \vdots & \ddots & \vdots \\[4pt]
\sfrac1{\sqrt{n\cdot\smash{(n{+}1)}}} & \sfrac1{\sqrt{n\cdot\smash{(n{+}1)}}} 
& \sfrac1{\sqrt{n\cdot\smash{(n{+}1)}}} & \ldots 
& \sfrac{-n}{\sqrt{n\cdot\smash{(n{+}1)}}} \\[6pt]
\sfrac1{\sqrt{n{+}1}} & \sfrac1{\sqrt{n{+}1}} & \sfrac1{\sqrt{n{+}1}} 
& \ldots & \sfrac1{\sqrt{n{+}1}}
\end{matrix}\right) 
\left(\begin{matrix} 
x_1 \\[6pt] x_2 \\[4pt] \vdots \\[4pt] x_n \\[6pt] x_{n+1} \end{matrix}\right)
\ ,
\ee
where the center of mass $y_{n+1}\equiv y_0=\sfrac1{\sqrt{n+1}}X$ has been 
added again.
Any global SO$(n)$ rotation of the $y_i$ yields an equivalent description.
The relative-coordinate parametrization of our $n{+}1$-particle model in terms
of $y_i$ offers a second interpretation, by reading the $y_i$ as the 
{\it absolute\/} coordinates of an $n$-particle system {\it without\/} 
translation invariance. By a slight abuse of notation, we denote
\be
U^\perp\bigl(x^\perp(y)\bigr)=U(y) \und F^\perp\bigl(x^\perp(y)\bigr)=F(y)\ .
\ee
Surely, it is also possible to `oxidize' an $n$-particle system without
translation invariance to a translation-invariant $n{+}1$-particle system,
by adding a $y_0$ coordinate and embedding into $\R^{n+1}$ via~(\ref{embed}).
Since the WDVV equation trivializes for $n\le2$, it is relatively easy to 
write down translation-non-invariant two-particle models or 
translation-invariant three-particle models. In fact, there is a functional
freedom in the choice~\cite{klp}. Therefore, in this paper we concentrate on
the nontrivial cases of $n\ge3$.

All known WDVV solutions are of the form~\footnote{
We disregard here the possibility of `radial' terms, where the argument of $K$
is $\sqrt{\sum_i x_i^2}$ or $\sqrt{\sum_{i<j}(x_i{-}x_j)^2}$~\cite{glp2,lp}.}
\be \label{Fansatz}
F(x) \= \sum_\beta f_\beta\,K(\beta{\cdot}x) \qquad\textrm{with}\qquad
f_\beta\in\R \und \beta{\cdot}x=\beta^i x_i\ ,
\ee
where the sum runs over a collection $\{\beta\}$ of $p$ non-parallel 
(co)vectors, and the function $K$ is universal up to a quadratic polynomial,
\bea \label{K}
K'''(z)=-\frac1z \quad&\Rightarrow&\quad 
K(z)=-\sfrac14 z^2\ln z^2
\qquad\qquad\quad\ \textrm{in the rational case}\ , \\[4pt] \label{Ktri}
K'''(z)=-\cot z  \quad&\Rightarrow&\quad 
K(z)=-\sfrac14\textrm{Li}_3(\ep^{2\im z})+\sfrac{\im}{6}z^3
\qquad\textrm{in the trigonometric case}\ , \\[4pt]
K'''(z)=-\frac{\vartheta^\prime_1(\frac{z}{\pi}|\tau)}
{\pi\,\vartheta_1(\frac{z}{\pi}|\tau)} \quad&\Rightarrow&\quad
K(z)=-\sfrac14{\cal L}i_3(\ep^{2\im z}|\tau)
\qquad\qquad\!\textrm{in the elliptic case}\ ,
\eea
where $\textrm{Li}_3$ is the trilogarithm and ${\cal L}i_3$ an elliptic
generalization~\cite{martini1}--\cite{strachan}.
For the prepotential~$U$ we make an ansatz which matches the form 
of~$F$,\footnote{
This may be very restrictive, 
as the solutions found in~\cite{klp} demonstrate.}
\be \label{Uansatz}
U(x) \= \sum_\beta u_\beta\,L(\beta{\cdot}x) \qquad\textrm{with}\qquad
u_\beta\in\R \und L'(z)=-K'''(z)\ ,
\ee
\be \label{L}
\textrm{thus}\qquad\qquad
L_{\textrm{rat}}=\sfrac12\ln z^2\ ,\qquad
L_{\textrm{tri}}=\sfrac12\ln\sin^2\!z\ ,\qquad
L_{\textrm{ell}}=\ln\vartheta_1(\sfrac{z}{\pi}|\tau)\ .
\qquad\qquad{}
\ee
Note that not all vectors from $\{\beta\}$ need to appear in $F$ or $U$,
because some $f_\beta$ or $u_\beta$ may vanish.
For illustration, we have given the `universal functions' $K$ and $L$ 
also for the trigonometric and elliptic models. However,
the normalization conditions~(\ref{conD}) imposed by conformal invariance
can only be satisfied in the rational case, and this is the only one treated
in this paper. Then, the normalizations~(\ref{conD}) translate into
simple conditions for the coefficients $u_\beta$ and $f_\beta$,
\be \label{norm}
\sum_\beta u_\beta \= 2\alpha \und 
\sum_\beta f_\beta\;\beta_j\beta_k \= (1{+}2\alpha)\,\delta_{jk}
\qquad\Rightarrow\qquad 
\sum_\beta \beta^2\,f_\beta \= (1{+}2\alpha)\,n \ ,
\ee
and the bosonic potential becomes
\be
V(x) \= \sfrac{g^2}{8}\sum_{\beta,\gamma} u_\beta u_\gamma\,
\frac{\beta\cdot\gamma}{\beta{\cdot}x\ \gamma{\cdot}x}\ .
\ee
We can actually employ the Killing-type equation~(\ref{eq1}) to solve for
$u_\beta$ in terms of~$f_\beta$.
When inserting the forms (\ref{Fansatz}) and~(\ref{Uansatz}) into~(\ref{eq1}), 
the vanishing of each double pole $(\beta{\cdot}x)^{-2}$ yields
\be \label{doublepole}
u_\beta(u_\beta{+}1)\=\beta^2 f_\beta\,u_\beta \qquad\Rightarrow\qquad
u_\beta=0 \qquad\textrm{or}\qquad
u_\beta=\beta^2 f_\beta-1
\ee
for each (co)vector~$\beta$, with $\beta^2\equiv\beta{\cdot}\beta$.
Inserting this into the `sum rule' (\ref{norm}) for~$u_\beta$, we obtain
a second necessary condition for $\{f_\beta\}$, namely
\be \label{norm2}
\sum_\beta\delta_\beta\,(\beta^2 f_\beta-1) \= 2\alpha 
\qquad\textrm{with}\qquad \delta_\beta\in\{0,1\}\ .
\ee
It restricts the $F$~solutions to those which may admit a $U$~solution as well.
However, by no means it guarantees that the single-pole terms in~(\ref{eq1})
work out as well.

Since we are only interested in permutation-invariant models, we demand
that the collection~$\{\beta\}$ of $p$~(co)vectors is closed under permutations
and that the coefficients $f_\beta$ and $u_\beta$ depend only on the orbit
$[\beta]$ of $\beta$. This suggests the notation (with square brackets)
\be
K_{n+1}[\beta{\cdot}x] \ :=\ \sum_{\pi} K\bigl(\pi(\beta){\cdot}x\bigr)
\und K_n[\beta{\cdot}y] \ :=\ \sum_{\pi} K\bigl(\pi(\beta){\cdot}y\bigr)
\ee
where the index indicates the particle number, and sum runs over all 
permutations which alter~$\pm\beta$. To indicate a particular orbit in
the square-bracket argument, we insert a typical representative and omit
the coordinate labels, e.g.
\be
K_3[3y{-}y{-}y]\ :=\ 
K(3y_1{-}y_2{-}y_3)+K(3y_2{-}y_3{-}y_1)+K(3y_3{-}y_1{-}y_2)\ ,
\ee
and likewise for the function $L$.

Finally, we comment on our solution strategy for the `structure equations'
(\ref{integr})--(\ref{eq1}). Guided by the known WDVV 
solutions~\cite{feives1,feives2}, we select a (co)vector 
collection~$\{\beta\}$, which gives us $F$ and~$U$ via (\ref{Fansatz})
and~(\ref{Uansatz}), with undetermined coefficients $f_\beta$ and~$u_\beta$.
Importing from the literature a particular solution for $\{f_\beta\}$, 
one remains with algebraic relations for $\{u_\beta\}$ which are very 
intricate, however.
Therefore, it is preferable not to start with some solution~$F$, but
to pick a structure only for~$U$ via~(\ref{Uansatz}) and then to regard
the WDVV and Killing-type equations (\ref{wdvv}) and~(\ref{eq1}) as
algebraic equations for the functions~$F_{ijk}$, disregarding their
integrability condition for the moment. Beyond three particles, the
(algebraic) WDVV equation~(\ref{wdvv}) becomes rather involved. Therefore,
as a detour, we first solve a simpler (linear) equation which follows from 
it and~(\ref{eq1}), namely~\footnote{
We thank A.~Galajinsky for a similar suggestion.}
\be\label{wdvv2}
(U_{ij}-U_iU_j)\,F_{jk\ell}\ -\ (U_{kj}-U_kU_j)\,F_{ji\ell}\= 0\ .
\ee
When $\{F_{ijk}\}$ has been constrained to obey this relation and 
also~(\ref{eq1}), it is much easier to completely solve the WDVV~equation.
In fact, given~$\{\beta\}$ and $\{u_\beta\}$, one can always construct 
a solution~$\{F_{ijk}\}$ in the form
\be
F_{ijk}(x)=-\sum_\beta\frac{f_{ijk}}{\beta{\cdot}x}\ .
\ee
The crucial point then is the integrability of those functions, i.e.
\be
\pa_{[i}F_{j]k\ell}=0 \qquad\Leftrightarrow\qquad
f_{jk\ell}=\sum_\beta f_\beta\,\beta_j\beta_k\beta_\ell\ ,
\ee
see (\ref{integr}) or~(\ref{integr1}).
It is a very restrictive requirement, which in many cases completely rules
out any solution~$\{u_\beta,f_\beta\}$. In other cases, it removes
any freedom in these coefficients (coming from WDVV-solution moduli)
and may even fix the value of the parameter~$\alpha$.

\section{$A$-type models}

\noindent
The simplest and most symmetric WDVV solutions take 
(the positive part of) the $A_n$ root system for $\{\beta\}$. 
The canonical representation lives in $\R^{n+1}$ in the hyperplane
orthogonal to~$\rho$,
\be \label{FAn}
F^\perp(x) \= \sfrac{1+2\alpha}{n+1}\,K_{n+1}[x{-}x]\ .
\ee
Unfortunately, for $n>3$ we could not find a $U$~solution, except when
\be \label{FUAn}
2\alpha=n: \qquad
F^\perp(x) \= K_{n+1}[x{-}x] \und
U^\perp(x) \= \sum_{i=1}^n L(x_i{-}x_{n+1})\ ,
\ee
which is not fully permutation invariant however.

It is known~\cite{chaves} that (\ref{FAn}) is a special point in 
an $n$-parameter family of WDVV solutions based on particular
deformations of the $A_n$ root system. These deformations also break
the permutation invariance of~$F^\perp$, but there exists a
one-parameter subfamily which is permutation symmetric in the
{\it relative\/} coordinates~$y_i$. Therefore, let us reduce the
description to~$\R^n$ and search for translation non-invariant
$n$-particle solutions. With the abbreviations \ $Y=\sum_iy_i$ \ and \
$\delta^2=\sfrac{n+1}{1+nt}$, the WDVV~solution reads
\be \label{FAnt}
F_t(y) \= \sfrac{1+2\alpha}{n+1} \Bigl\{ (1{-}t)\,K_n\bigl[y-y\bigr]
\ +\ \sfrac{1+nt}{n^2}\, K_n\bigl[(ny-(1{+}\delta)Y] \Bigr\}\ ,
\ee
where the first term contains an $A_{n-1}$ subsystem and the remaining
$n$ roots are deformed by changing their component in the direction
of the subsystems center-of-mass vector $\rho=(1,1,\ldots,1)$.
As $t\in[-\sfrac1n,\infty]$, we cover the four cases
\begin{center}
\begin{tabular}{cccc}
{}\quad\qquad$t=-\sfrac1n$\quad\qquad{} & {}\quad$t=0$\qquad{} & 
\qquad{}$t=1$\qquad{} & {}\qquad$t=\infty$\qquad{} \\[4pt]
$\delta=\infty$ & $\delta=\sqrt{n{+}1}$ & $\delta=1$ & $\delta=0$ \\[4pt]
$A_{n-1}\oplus A_1$ & $A_n$ & $A_1^n$ & $A_{n-1}\oplus\underline{n}$
\end{tabular}
\end{center}
where $\underline{n}$ denotes the fundamental (quark) weights
of $A_{n-1}$. 

Which of these $F$ backgrounds admit a $U$~solution? 
Permutation invariance applied to~(\ref{norm2}) leaves us with three options,
corresponding to the choices of~$\delta_\beta$:
$U$ may contain either only the first type of (co)vectors from~(\ref{FAnt}),
or only the second type of (co)vectors, or both.
In each case (\ref{norm}) yields an equation of the form~$h_n(t,\alpha)=0$.
Numerical analysis reveals that only the second option fully works out,
\be
U(y)\=u\,L_n\bigl[(ny-(1{+}\delta)Y\bigr]
\qquad\textrm{with}\quad u=\sfrac{1-t}{1+nt}=\sfrac{2\alpha}{n}\ .
\ee
The corresponding condition from~(\ref{norm2}) is
\be \label{Arel}
(1{+}2\alpha)(nt+1)=n+1 \qquad\Leftrightarrow\qquad
t=\frac1n\frac{n{-}2\alpha}{1{+}2\alpha} \qquad\Leftrightarrow\qquad
1{+}2\alpha=\frac{n{+}1}{nt{+}1}\ .
\ee
For $t{=}0$ ($\alpha{=}\frac{n}2$) this yields an undeformed $A_n$~solution
where in $U$ only $n$ of the roots appear:
\be
\begin{aligned}
F_0(y) &\= K_n\bigl[y-y\bigr] \ +\ 
\sfrac{1}{n^2}\, K_n\bigl[ (ny-(1{+}\sqrt{n{+}1})Y \bigr]\ ,
\\[4pt]
U_0(y) &\= L_n\bigl[ (ny-(1{+}\sqrt{n{+}1})Y \bigr]\ ,
\end{aligned}
\ee
which is the reduced form of~(\ref{FUAn}).
For $n{=}3$ it simplifies to
\be \label{A3sol}
F_0(y) \= K_3[y-y] \ +\ K_3[y+y] \und U_0(y)\= L_3[y+y]\ .
\ee

The limit $t\to\infty$ ($\delta\to0$) deserves special attention, 
because the induced metric~$G$ degenerates. For this reason, this boundary 
of the WDVV solution space is normally excluded in the mathematical literature
(see, however, \cite{strachan}). 
Also, (\ref{FAnt}) tells us that we need to tune
\be
1{+}2\alpha=0: \qquad
F_\infty(y) \= -\sfrac1n K_n[y-y]\ +\ \sfrac{1}{n^2}\,K_n[ny-Y] \und
U_\infty(y) \= -\sfrac1n\,L_n[ny-Y]\ ,
\ee
where the scale of~$F_\infty$ is determined via~(\ref{norm2}).
In this limit, our deformed root system fits in the hyperplane 
orthogonal to~$\rho$, thus we recover translation invariance! 
Therefore, only for $osp(4|2)$ symmetry we have an $n$-particle solution which
meets all physical requirements. Its bosonic potential
\be
V(y)\=\frac{g^2}8\,\Bigl\{ \sum_i \frac1{(ny_i{-}Y)^2}\ -\
\frac{1}{n} \Bigl( \sum_i \frac1{(ny_i{-}Y)} \Bigr)^2 \Bigr\} 
\ee
however, is not
of the Calogero type, because only the fundamental weights contribute to it.

Due to the isometry $A_3\simeq D_3$,
the reduction of the $A_4(\infty)$ solution to $\R^3$ remains permutation
invariant (see the following section),
\bea
&& F_\infty(y) \= -\sfrac14 K_4[y{-}y]\ +\ \sfrac{1}{16} K_4[3y{-}y{-}y{-}y]
\quad\longrightarrow\qquad 
-\sfrac14 K_3[y{\pm}y]\ +\ \sfrac14 K_3[y{\pm}y{\pm}y]\ , \\[4pt]
&& U_\infty(y) \= -\sfrac14 L_4[3y{-}y{-}y{-}y]
\qquad\qquad\qquad\quad\;\;\,\longrightarrow\qquad 
-\sfrac14 L_3[y{\pm}y{\pm}y]\ .
\eea
One may wonder whether further solutions can be produced by admitting other
weights to the ans\"atze (\ref{Fansatz}) and~(\ref{Uansatz}). This is not 
the case, except for $n\le4$, where accidents happen due to the existence 
of~$F_4$ and the isometries $A_3\simeq D_3$ and 
$A_3\oplus\underline{6}\simeq B_3$.
These cases are more naturally described in the following sections.

\section{$BCD$-type models}

\noindent
The $B_n$, $C_n$ and $D_n$ root systems do not yield
permutation symmetric models in~$\R^{n+1}$, but are naturally formulated
in the {\it relative\/} coordinates~$y_i\in\R^n$. Therefore, we consider the
reduced description, which trades translation invariance for permutation 
invariance. The WDVV equation~(\ref{wdvv}) has an $n$-parameter family
of solutions based on deformed $BCD_n$ roots~\cite{chaves}, 
\be \label{FBCDfull}
F_{t,\vec s}(y)\=\frac{1{+}2\alpha}{2(s^2{-}\delta^2)} \biggl\{ \,
\sum_{i<j} K_n\bigl(s_jy_i{\pm}s_iy_j\bigr)\ +\ 
2\sum_i K_n\bigl(\sqrt{\smash{s_i^2{-}\delta^2}\phantom{|}}y_i\bigr)
\biggr\} \qquad\textrm{with}\quad \delta^2=\frac{1{-}nt}{1{-}t}\ ,
\ee
where $t,s_i\in\R$ for $i=1,\ldots,n$, and one parameter is redundant.
To retain permutation symmetry, we keep the roots undeformed, $s_i=1$, but 
allow $t$ to vary, so that
\be \label{FBCD}
F_t(y)\=\lim_{s_i\to1}F_{t,\vec s}(y)\=(1{+}2\alpha) \Bigl\{
\sfrac{1-t}{2n-2}\,K_n[y{\pm}y]\ +\ t\,K_n[y]\Bigr\}\ .
\ee
By changing the parameter~$t$, we reach four special cases in $BCD_n(t)$:
\begin{center}
\begin{tabular}{cccc}
{}\qquad$t=0$\qquad{} & {}\quad$t=\sfrac1{2n-1}$\quad{} & 
{}\quad$t=\sfrac2{n+1}$\quad{} & {}\qquad$t=1$\qquad{} \\[4pt]
$D_n$ & $B_n$ & $C_n$ & $A_1^n$
\end{tabular}
\end{center}

Similarly to the $A_n(t)$ deformation, it turns out that most $U$~solutions
carry only the short roots. For this case, (\ref{norm2}) yields the 
relation~\footnote{
It is remarkable that this $BCD_n(t)$ relations is obtained from the
$A_n(t)$ relation~(\ref{Arel}) by $n\to-n$.}
\be \label{Brel}
(1{+}2\alpha)(nt-1)=n-1 \qquad\Leftrightarrow\qquad
t=\frac1n\frac{n{+}2\alpha}{1{+}2\alpha} \qquad\Leftrightarrow\qquad
1{+}2\alpha=\frac{n{-}1}{nt{-}1}\ ,
\ee
which may be used to fix $t=t(\alpha)$ in the solution
\be \label{BCDsol}
F(y)\=\sfrac1n\Bigl\{ \alpha\,K_n[y{\pm}y]\ +\ (n{+}2\alpha)\,K_n[y]\Bigr\}
\und
U(y)\=\sfrac{2\alpha}{n}\,L_n[y]\ .
\ee
Again, only $n$ of the roots appear in~$U$.
Of course, we may instead fix $\alpha=\alpha(t)$ and read off solutions for
\bea
B_n: \quad 1{+}2\alpha=1{-}2n &\quad\Rightarrow\quad&
F(y)\=-K_n[y{\pm}y]-K_n[y] \und U(y)\=-2L_n[y] \ ,\\[4pt]
C_n: \quad 1{+}2\alpha=1{+}n \ \, &\quad\Rightarrow\quad&
F(y)\=\sfrac12 K_n[y{\pm}y]+2\,K_n[y] \!\!\!\und U(y)\=L_n[y] \ ,\\[4pt]
D_n: \quad 1{+}2\alpha=1{-}n \ \, &\quad\Rightarrow\quad&
F(y)\=-\sfrac12 K_n[y{\pm}y] \qquad\quad\,\und U(y)\=-L_n[y] \ ,
\eea
where the $D_n$ case employs the vector weights~$(v)$ in~$U$.

For $n{=}4$ the triality automorphism~$T$ of~$D_4$ generates additional 
solutions from the ones above, via
\be \label{triality}
\left(\begin{matrix}
y_1 \\[2pt] y_2 \\[2pt] y_3 \\[2pt] y_0 \end{matrix}\right)
\quad\buildrel{T}\over{\longmapsto}\quad \frac12
\left(\begin{matrix}
1 & -1   & -1   & -1   \\[2pt]
1 & -1   & \ph1 & \ph1 \\[2pt]
1 & \ph1 & -1   & \ph1 \\[2pt]
1 & \ph1 & \ph1 & -1 
\end{matrix}\right)
\left(\begin{matrix}
y_1 \\[2pt] y_2 \\[2pt] y_3 \\[2pt] y_4 \end{matrix}\right) 
\qquad\textrm{with}\quad T^3=-\unity\ ,
\ee
under which the vector~$(v)$, spinor~$(s)$ and conjugate spinor~$(c)$
representations get cycled around. 
This allows us to find a couple of further solutions, in which $U$ carries 
incomplete Weyl orbits of roots. 
Since the fully deformed WDVV~solution~(\ref{FBCDfull}) breaks the Weyl 
symmetry of the $B_n$~system, we may ignore it even in the limit $s_i\to1$.
Making use of this freedom, we single out $y_1$ and find the $n{=}4$ solutions
\bea
1{+}2\alpha=6: && F(y)\=K_4[y{\pm}y] \qquad\quad\ \ \und 
U(y)\=L_4[y_1{\pm}y]-L_4(y_1) \ ,\\[4pt]
1{+}2\alpha=7: && F(y)\=K_4[y{\pm}y]+K_4[y] \und
U(y)\=L_4[y_1{\pm}y] \ ,
\eea
where $[y_1{\pm}y]$ stands for $\{y_1{\pm}y_2,y_1{\pm}y_3,y_1{\pm}y_4\}$.
Acting with a triality transformation~(\ref{triality}), we obtain two 
permutation-invariant solutions with $p{=}13$ and $p{=}16$, respectively:
\bea
1{+}2\alpha=6: && F(y)=K_4[y{\pm}y] \qquad\qquad\qquad\qquad\ \ \,\und 
U(y)=L_4[y{+}y]-L_4(Y) \ ,\\[4pt]
1{+}2\alpha=7: && F(y)=K_4[y{\pm}y]+\sfrac14 K_4[y{+}y{\pm}y{\pm}y]_+ \und
U(y)=L_4[y{+}y] \ ,
\eea
where $[\ldots]_+$ indicates an even number of minuses in the bracket.

The case of $n{=}3$ is also special. 
For $\alpha{=}1$ it admits an isolated additional solution,
\be \label{B3sol}
F(y)\=\sfrac23\,K_3[y{\pm}y]\ +\ \sfrac13\,K_3[y] 
\und U(y)\=\sfrac13\,L_3[y{\pm}y]\ .
\ee
Due to the isomorphy $D_3\simeq A_3$, the oxidation of the $BCD_3$ models
to $\R^4$ can be made permutation invariant, by reading the $D_3$ weights
as $A_3$ weights. With an SO(3) rotation built from the triality
map~(\ref{triality}), the explicit coordinate relation reads
($y_4\equiv y_0$)
\be \label{embed34}
y_i \= T_{ij}\,x_j \ ,
\ee
and we obtain the following translation table,
\begin{center}
\begin{tabular}{lcccccc}
representation & \underline{4} & \underline{6} & \underline{15} & 
\underline{45} & \underline{64} & $\cdots$ \\[2pt]
length of orbit   & 4 & 3 & 6 & 12 & 12 & $\cdots$ \\[2pt]
argument $[\beta{\cdot}y]$ & 
$[y{\pm}y{\pm}y]$ & $[y]$ & $[y{\pm}y]$ & 
$[2y{\pm}y{\pm}y]$ & $[2y{\pm}y]$ & $\cdots$ \\[2pt]
argument $[\beta{\cdot}x]$ & 
$\sfrac12[3x{-}x{-}x{-}x]$ & $\sfrac12[x{+}x{-}x{-}x]$ & $[x{-}x]$ & 
$[2x{-}x{-}x]$ & $\sfrac12[3x{-}3x{+}x{-}x]$ & $\cdots$
\end{tabular}
\end{center}
By oxidizing the $\underline{15}$ and $\underline{6}$ weights and using
$K_n[\lambda\beta{\cdot}x]\simeq\lambda^2 K_n[\beta{\cdot}x]$ and
$L_n[\lambda\beta{\cdot}x]\simeq L_n[\beta{\cdot}x]$, we can formulate
translation-invariant $BCD_3$ models,
\bea
\textrm{any $\alpha$:} &&
F^\perp(x)=\sfrac{\alpha}{3}K_4[x{-}x]-\sfrac{3+2\alpha}{12}K_4[x{+}x{-}x{-}x]
\quad\textrm{and}\quad \label{BCD3ox}
U^\perp(x)=\sfrac{2\alpha}{3}L_4[x{+}x{-}x{-}x] \ ,\\[4pt]
\alpha{=}1: && 
F^\perp(x)=\sfrac23 K_4[x{-}x]+\sfrac{1}{12}K_4[x{+}x{-}x{-}x] 
\!\ \qquad\textrm{and}\quad \label{B3ox}
U^\perp(x)=\sfrac13 L_4[x{-}x] \ .
\eea
Note that only the last model, which is invariant under 
$D(2,1;1)\simeq osp(4|2)$, gives rise to a Calogero potential~$V$.
Some $D_n$ spinor weights occur in further solutions, but these are
more naturally obtained within the $F$-type models, which derive from $E_8$
and are discussed next.

\section{$EF$-type models}

\noindent
Further WDVV solutions are based on the root systems of the exceptional
simple Lie algebras. For rank $n>2$, they can all be obtained by reducing
the $E_8$ system in particular ways. Also exceptional deformed root systems
appear in this way, as projections of $E_n$ or $F_4(t)$ along some parabolic
subgroup~\cite{feives1}.\footnote{
Non-crystallographic Coxeter root systems 
do not produce permutation-invariant systems.}
Among this variety, we restrict ourselves to permutation-symmetric models
for physical reasons. This leaves us with the following possibilities.

With $p{=}120$ (co)vectors the $E_8$ system is the largest exceptional one,
\be
F(y)_\pm\=\sfrac{1+2\alpha}{30} \Bigl\{ K_8[y{\pm}y]\ +\
\sfrac14 K_8[y{+}y{+}y{+}y{\pm}y{\pm}y{\pm}y{\pm}y]_\pm \Bigr\} \ ,
\ee
where the `$\pm$' subscript indicates an even or odd number of minuses.
The two solutions are related by the standard spinor helicity flip.
The $E_7$ system with $p{=}63$ is more naturally formulated as a 
translation-invariant eight-particle model,
\be
F^\perp(x)\=\sfrac{1+2\alpha}{18} \Bigl\{ K_8[x-x]\ +\
\sfrac14 K_8[x{+}x{+}x{+}x{-}x{-}x{-}x{-}x] \Bigr\} \ .
\ee
$E_6$ is not permutation symmetric.
Neither case allows for a $U$~solution, so no such models exist.

Therefore, we pass to the $F_n$ series as defined in~\cite{feives1} by
the projection of the $E_8$ system along its $D_{8-n}$ subgroup,
for $n=3,4,5,6$.
First, $F_6\simeq(D_8,A_1^2)$ with $p{=}68$ yields
\bea
&& F(y)\=\sfrac{1+2\alpha}{30} \Bigl\{ K_6[y{\pm}y]\ +\ 4 K_6[y]\ +\
\sfrac12 K_6[y{+}y{+}y{\pm}y{\pm}y{\pm}y] \Bigr\} \ , \\[4pt]
&& 2\alpha{+}1=15: \quad
U(y)_\pm\=L_6[y]\ +\ \sfrac12 L_6[y{+}y{+}y{\pm}y{\pm}y{\pm}y]_\pm \ ,
\eea
with the same notation as above.
Second, $F_5\simeq(E_8,A_3)$ has $p{=}41$ and produces
\bea
&& F(y)\=\sfrac{1+2\alpha}{30} \Bigl\{ K_5[y{\pm}y]\ +\ 6 K_5[y]\ +\
K_5[y{+}y{+}y{\pm}y{\pm}y] \Bigr\} \ , \quad\ {} \\[4pt]
&& 2\alpha{+}1=\sfrac{15}{2}: \quad
U(y)\=\sfrac12 L_5[y]\ +\ \sfrac14 L_5[y{+}y{+}y{\pm}y{\pm}y] \ .
\eea

In the next reduction step, we meet $(E_8,D_4)\simeq F_4$ with $p{=}24$
(co)vectors, which as a Lie algebra with two Weyl orbits allows for a
one-parameter deformation~$F_4(t)$,
\be
F_t(y)\=(1{+}2\alpha) \Bigl\{ \sfrac{1-t}{6} K_4[y{\pm}y]\ +\ 
\sfrac{t}{3} K_4[y]\ +\ \sfrac{t}{12} K_4[y{+}y{\pm}y{\pm}y] \Bigr\}\ .
\ee
It is invariant under the exchange of its two $D_4$ subsystems while 
$t\to1{-}t$. Special values of~$t$ are
\begin{center}
\begin{tabular}{ccc}
{}\qquad$t=0$\qquad{} & {}\quad$t=\sfrac13$\quad{} & {}\qquad$t=1$\qquad{}
\\[4pt]
$D_4$ & $F_4$ & $D'_4$ \\[4pt]
$p{=}12$ & $p{=}24$ & $p{=}12$
\end{tabular}
\end{center}
where the $D_4$ and $D'_4$ systems are formed by the long and short roots
of~$F_4$, respectively.
The three types of roots allow for more options in~$U$ than was the case
in the $A_n$ or $BCD_n$ models. On the corresponding curves $h_n(t,\alpha)=0$,
howver, only isolated $U$~solutions occur:\footnote{
We do expect families of solutions whose generic members, however, will not
be permutation invariant. The corresponding curves are \
$(1{+}2\alpha)(4t{-}3)=9$, \ $(1{+}2\alpha)(4t{-}3)=-11$, \
$(1{+}2\alpha)(4t{-}1)=11$ \ and \ $(1{+}2\alpha)(8t{-}3)=21$, respectively.}
\bea
1{+}2\alpha=-3 \quad\&\quad t=0\ : &&
U(y)\= -L_4[y] \quad\textrm{or}\quad -L_4[y{+}y{\pm}y{\pm}y]_\pm\ , \\[4pt]
1{+}2\alpha=+5 \quad\&\quad t=\sfrac15\ : &&
U(y)\=\sfrac13 L_4[y{\pm}y] \ , \\[4pt]
1{+}2\alpha=+5 \quad\&\quad t=\sfrac45\ : &&
U(y)\=\sfrac13 L_4[y] + \sfrac13 L_4[y{+}y{\pm}y{\pm}y] \ , \\[4pt]
1{+}2\alpha=+9 \quad\&\quad t=\sfrac23\ : &&
U(y)\= L_4[y] + L_4[y{+}y{\pm}y{\pm}y]_\pm \quad\textrm{or}\quad
L_4[y{+}y{\pm}y{\pm}y]\ .
\eea
The three solutions in the first line and also in the fourth one are related 
by triality (the very first solution occurred already under $D_4$).
The $D_4{\leftrightarrow}D'_4$ flip applied to lines one or four yields
permutation non-invariant configurations (which we ignore here), 
but relates the solutions in the second and third lines, which are
triality invariant by themselves.

The final reduction yields the $F_3(t)$ family with $p{=}13$ and
\be
F_t(y)\=(1{+}2\alpha) \Bigl\{ \sfrac{1-t}{6} K_3[y{\pm}y]\ +\ \sfrac13 K_3[y]
\ +\ \sfrac{t}{6} K_3[y{\pm}y{\pm}y] \Bigr\} \ ,
\ee
which connects the $BCD_3(t)$ family to the $A_4(t)$ one,
\begin{center}
\begin{tabular}{ccc}
{}\qquad$t=0$\qquad{} & {}\qquad$t=1$\qquad{} & {}\qquad$t=\infty$\qquad{} 
\\[4pt]
$BCD_3(\sfrac13)$ & $D(2,1;\alpha)$ & $A_4(\infty)=A_3\oplus\underline{4}$
\\[4pt]
$p{=}9$ & $p{=}7$ & $p{=}10$
\end{tabular}
\end{center}
In this case, $U$~solutions occur in {\it two\/} subfamilies and one
sporadic case:
\bea
&& (1{+}2\alpha)\,t=3: \qquad\qquad\qquad \label{F3asol}
U(y)\=\sfrac{2\alpha-2}{3} L_3[y]\ +\ \sfrac12 L_3[y{\pm}y{\pm}y] \ ,\\[4pt]
&& (1{+}2\alpha)(2t{-}1)=3: \qquad\quad\ \; \label{F3bsol}
U(y)\=\sfrac{\alpha}{2} L_3[y{\pm}y{\pm}y] \ ,\\[4pt]
&& t=\sfrac15 \quad\&\quad 1{+}2\alpha=5: \qquad\; \label{F3csol}
U(y)\=\sfrac13 L_3[y{\pm}y]\ +\ \sfrac23 L_3[y] \ .
\eea
Indeed, at $(t{=}0\,,1{+}2\alpha{=}\infty)$ the first family matches to the
$BCD_3(\sfrac13)$ solution, and at $(t{=}\infty\,,1{+}2\alpha{=}0)$ the second 
family agrees with the $A_4(\infty)$ one.
Employing the embedding~(\ref{embed34}) and the subsequent translation table,
we can oxidize these systems to translation-invariant four-particle 
models,\footnote{
For $\alpha{=}0$, the first one yields a $C{=}0$ four-particle solution 
to~(\ref{Weq}), \ 
$W=\ep^{-U}=\prod(x{+}x{-}x{-}x)^{2/3}\prod(3x{-}x{-}x{-}x)^{-1/2}$ 
in obvious notation, which was missed in~\cite{klp}.}
\bea
&& \begin{cases} 
F^\perp(x)\=\sfrac{\alpha-1}{3}K_4[x{-}x]\ +\
\sfrac{1+2\alpha}{12}K_4[x{+}x{-}x{-}x]\ +\
\sfrac18 K_4[3x{-}x{-}x{-}x] \\[2pt]
U^\perp(x)\=\sfrac{2\alpha-2}{3}L_4[x{+}x{-}x{-}x]\ +\
\sfrac12 L_4[3x{-}x{-}x{-}x]
\end{cases} \ ,\label{F3aox} \\[4pt]
&& \begin{cases} 
F^\perp(x)\=\sfrac{\alpha-1}{6}K_4[x{-}x]\ +\
\sfrac{1+2\alpha}{12}K_4[x{+}x{-}x{-}x]\ +\
\sfrac{2+\alpha}{24}K_4[3x{-}x{-}x{-}x] \\[2pt]
U^\perp(x)\=\sfrac{\alpha}{2}L_4[3x{-}x{-}x{-}x] 
\end{cases} \ ,\label{F3box} \\[4pt]
&& \begin{cases}
F^\perp(x)\=\sfrac23 K_4[x{-}x]\ +\
\sfrac{5}{12}K_4[x{+}x{-}x{-}x]\ +\
\sfrac{1}{24}K_4[3x{-}x{-}x{-}x] \\[2pt]
U^\perp(x)\=\sfrac13 L_4[x{-}x]\ +\ 
\sfrac23 L_4[x{+}x{-}x{-}x]
\end{cases} \ .\label{F3cox}
\eea

\section{Non-Coxeter-type models}

\noindent
It is known that the root systems of some Lie superalgebras also
give rise to WDVV~solutions~\cite{feives1,serves}.
Of interest are one-parameter deformations of $AB(1,3)$ and $G(1,2)$
and a two-parameter deformation of~$D(2,1;\alpha)$. 
The $AB(1,3)$ family admits two inequivalent reductions to $n{=}3$,
one of which yields a permutation-symmetric solution with $p{=}10$:
\be 
F_t(y)=\sfrac{1{+}2\alpha}{27(t^2{+}1)} \Bigl\{ 
9 K_3[y{-}y] + K_3[ty{+}ty{-}2ty{+}2wY] + 
2 K_3[ty{+}ty{-}2ty{-}wY] + \sfrac92(t^2{-}1)K_3[Y] \Bigr\}
\ee
with \ $w^2=\sfrac14(t^2{+}3)$ \ for $t\in\R_+$.
At $t{=}1$ there exists a full solution for~$(F,U)$:
\be \label{AB3sol}
1{+}2\alpha=6: \qquad
F(y)\= K_3[y{\pm}y] + 2 K_3[y] \und
U(y)\= L_3[y{+}y] + L_3[y] - L_3[Y]\ .
\ee
We know of no other non-Coxeter-type permutation-invariant solutions.

\section{Conclusions}

\noindent
By adding to the particle coordinates and their superpartners
a single harmonic variable (parametrizing a two-sphere), 
we have overcome the technical barrier for constructing 
${\cal N}{=}4$ superconformal mechanics models with more than three particles.
The structure equations determining the two prepotentials $F$ and~$U$ 
admit simple solutions based on deformed root systems, for an arbitrary number 
of particles and for the superconformal symmetry algebra $D(2,1;\alpha)$ 
at any value of~$\alpha$.
We have restricted ourselves to permutation-invariant prepotentials and
performed a numerical survey of all permutation-symmetric (deformed) root
configurations, with and without translation invariance.

In each moduli space of WDVV~solutions~$F$ based on a deformed $A_n$ or $BCD_n$
root system, we have identified a permutation-invariant one-parameter~$(t)$
subfamily. It turns out that the solutions of the Killing-type equation
for the second prepotential~$U$ in the background of a given WDVV solution~$F$
live on a curve $h_n(t,\alpha)=0$ in the $(t,\alpha)$~plane. 
The $A_n(t{=}\infty)$ model,
built on the roots and fundamental weights of~$A_{n-1}$, is degenerate but 
distinguished by its translation invariance. Since $h_n(\infty,-\sfrac12)=0$,
this solution exists only for the $osp(4|2)$ case. Also, its bosonic potential
is not of Calogero-type. All other solutions lack translation invariance.
Of course, one may reinterpret their variables as {\it relative\/} particle
coordinates and add the center of mass to reclaim translation invariance,
but permutation symmetry will usually be lost in the new variables.

An exception occurs at $n{=}3$ because of the $A_3\simeq D_3$ isometry.
Inside the $F_3$ family (with parameter~$t$) of WDVV~solutions (a reduction 
of the $F_4$~family), we have identified two curves $h_3^{(1,2)}(t,\alpha)=0$
and one isolated point $(\hat{t},\hat{\alpha})$ for $U$~solutions. 
The corresponding models all lift to translation-invariant four-particle 
systems. For the $n{>}3$ exceptional root systems ($F_4$ and $E_n$ and 
reductions thereof) and also for some super root system (a reduction 
of $AB_4$), only sporadic solutions for particular values of~$\alpha$
and without translation invariance occur. We did not discuss the
$n{=}2$ systems, because (for $\alpha{=}0$) they have already been 
investigated thoroughly and are much less restrictive.
We have also constructed some solutions for the
trigonometric case, but not displayed them here.

Obviously lacking is a geometrical understanding of the `zoo' of solutions.
It would be nice to find {\it sufficient\/} conditions on~$\alpha$
or on $(t,\alpha)$ for the existence of $U$~solutions in a given
$F$~background. This may become more transparent if the requirement of
permutation symmetry is dropped, so that further $(F,U)$~solutions can be
revealed. Although this requirement is physically reasonable (and this only
for the full translation-invariant system), it is mathematically unnatural.
Perhaps a superspace reformulation of our models will shed more light on
this question.

\subsubsection*{Acknowledgments}

\noindent
We acknowledge support from a DFG grant, project No 436 RUS/113/669,
the RFBR grants 09-02-01209 and 09-02-91349 as well as a grant of the 
Heisenberg-Landau Program.
We are grateful to Misha Feigin, Anton Galajinsky, Konrad Schwerdtfeger
and Johannes Th\"urigen for help and fruitful discussions.
Sergey Krivonos thanks the Institut f\"ur Theoretische Physik at Leibniz
Universit\"at Hannover for hospitality.

\section*{Appendix}

\noindent
For an overview of the `zoo' of irreducible $n{=}3$ solutions,
we collect them all, namely (\ref{A3sol}), (\ref{BCDsol}), (\ref{B3sol}),
(\ref{F3asol})--(\ref{F3csol}) and (\ref{AB3sol}), in the following table,
\begin{center}
\begin{tabular}{|lc|ccccc|ccccc|}
\multicolumn{2}{c}{ } &
\multicolumn{5}{c}{
$\overbrace{\smash{\phantom{-----------------}}}^{
\textstyle\textrm{coefficients $f_{[\beta]}$ for $[\beta]=\ldots$}}$} &
\multicolumn{5}{c}{
$\overbrace{\smash{\phantom{-----------------}}}^{
\textstyle\textrm{coefficients $u_{[\beta]}$ for $[\beta]=\ldots$}}$} \\
\hline
system $\phantom{\Big|}$ & $\alpha$
& $[y{-}y]$ & $\ \;[y{+}y]\ \;$ & $\ [y]\ $ & $[y{\pm}y{\pm}y]$ & $[Y]$
& $[y{-}y]$ & $\ \;[y{+}y]\ \;$ & $\ [y]\ $ & $[y{\pm}y{\pm}y]$ & $[Y]\!$ \\
\hline
$A_3(0)\phantom{\Big|}$ & $\sfrac32$ & 1 & 1 & 0 & 0 & 0
& 0 & 1 & 0 & 0 & 0\\[4pt]
$A_4(\infty)$ & ${}\!\!\!{-}\sfrac12$ & $-\lambda$ & $-\lambda$ & 0
& $\lambda$ & 0 & 0 & 0 & 0 & $-\sfrac14$ & 0 \\[4pt]
$BCD_3$ & $\alpha$ & $\sfrac{\alpha}3$ & $\sfrac{\alpha}3$
& $\sfrac{2\alpha+3}3$ & 0 & 0 & 0 & 0 & $\sfrac{2\alpha}3$ & 0 & 0 \\[4pt]
$BCD_3(\sfrac59)\!\!\!{}$ & 1 & $\sfrac23$ & $\sfrac23$ & $\sfrac13$ & 0 & 0
& $\sfrac12$ & $\sfrac12$ & 0 & 0 & 0 \\[4pt]
$F_3^{(1)}$ & $\alpha$ & $\sfrac{\alpha-1}3$ & $\sfrac{\alpha-1}3$
& $\sfrac{2\alpha+1}3$ & $\sfrac12$ & 0 & 0 & 0 & $\sfrac{2\alpha-2}3$
& $\sfrac12$ & 0 \\[4pt]
$F_3^{(2)}$ & $\alpha$ & $\sfrac{\alpha-1}6$ & $\sfrac{\alpha-1}6$
& $\sfrac{2\alpha+1}3$ & $\sfrac{\alpha+2}6$ & 0 & 0 & 0 & 0
& $\sfrac{\alpha}2$ & 0 \\[4pt]
$F_3(\sfrac15)$ & 2 & $\sfrac23$ & $\sfrac23$ & $\sfrac53$ & $\sfrac16$ & 0
& $\sfrac13$ & $\sfrac13$ & $\sfrac23$ & 0 & 0 \\[4pt]
$AB_3(1)$ & $\sfrac52$ & 1 & 1 & 2 & 0 & 0 & 0 & 1 & 1 & 0 & $-1$\\[4pt]
\hline
\end{tabular}
\end{center}
Except for the first and last lines, all systems can be `oxidized' to
translation-invariant four-particle models, see (\ref{BCD3ox}), (\ref{B3ox})
and (\ref{F3aox})--(\ref{F3cox}).


\end{document}